\documentclass[cits]{PoS}

\usepackage[utf8]{inputenc}
\usepackage{amsmath}
\usepackage{dsfont}
\usepackage{subfig}
\usepackage{booktabs}
\usepackage{siunitx}

\graphicspath{{graphics/}}

\title{Numerical Stochastic Perturbation Theory \\ and the Gradient Flow}
\ShortTitle{NSPT and the Gradient Flow}

\author{\speaker{Mattia Dalla Brida}
	\thanks{Current address: NIC, DESY, Platanenallee 6, 15738 Zeuthen, Germany.}\\
        School of Mathematics, Trinity College Dublin, Dublin 2, Ireland\\
        E-mail: \email{mattia@maths.tcd.ie}}

\author{Dirk Hesse\\
        Universit\`a degli studi di Parma, Viale G.P. Usberti 7/a, 43100 Parma, Italy\\
        E-mail: \email{dirk.hesse@fis.unipr.it}}

\abstract{We study the Yang-Mills gradient flow using numerical stochastic
	  perturbation theory. As an application of the method we consider the 
	  recently proposed gradient flow coupling in the Schr\"odinger functional
	  for the pure SU(3) gauge theory.}

\FullConference{31st International Symposium on Lattice Field Theory - LATTICE 2013\\
		July 29 - August 3, 2013\\
		Mainz, Germany}

\begin{document}

\section{The gradient flow coupling}

In these Proceedings we focus on the pure SU(3) gauge theory. The Yang-Mills 
gradient flow $B_\mu(t,x)$, $t\geq0$, is then defined by the flow equation 
(see~\cite{Luscher:2010iy} for an introduction)
\begin{eqnarray}
  &&\hspace{-0.5em}\partial_tB_{\mu}=D_{\nu}G_{\nu\mu},\quad B_\mu|_{t=0} = A_\mu,
  \label{eq:GF}\\[2.0ex]
  &&\hspace{-0.5em}G_{\mu\nu}=\partial_{\mu}B_{\nu}-\partial_{\nu}B_{\mu}
  +[B_{\mu},B_{\nu}],
  \qquad
  D_{\mu}=\partial_{\mu}+[B_{\mu},\,\cdot\,],
\end{eqnarray}
where $A_\mu(x)$ is the fundamental gauge potential integrated over in the functional 
integral. A key feature of the flow is that correlation functions of the field 
$B_\mu(t,x)$ are automatically finite for flow times $t>0$  once the bare parameters
of the theory have been renormalized~\cite{Luscher:2011bx}. Such correlation functions
are thus well-defined observables and provide appealing probes for the properties
of the theory. One particularly interesting application of the flow is the non-perturbative
determination of the running coupling. Through observables at positive flow time
one can naturally define a renormalized coupling~\cite{Luscher:2010iy} and compute
its scale evolution using step scaling~\cite{Luscher:1991wu}.

Consider for example the flow energy density,
\begin{equation}
\label{eq:Eoft}
  \big\langle E(t) \big\rangle  =
  -\frac{1}{2}\,\big\langle{\rm Tr}\{G_{\mu\nu}(t)G_{\mu\nu}(t)\}\big\rangle.
\end{equation}
As shown in~\cite{Luscher:2010iy}, a renormalized coupling can be defined in 
infinite volume as
\begin{equation}
 \overline{g}^{2}(\mu) =  \frac{(4\pi)^2}{3} \big\langle t^2  E(t) \big\rangle,\quad
  \mu = 1/\sqrt{8t},
 \end{equation}
where the renormalization scale $\mu$ is set by the flow time. In a finite volume
one can analogously define a running coupling by scaling the renormalization scale
$\mu$ and all other dimensionful quantities in the system in fixed proportions to
the finite spatial extent. Such a definition has been studied in a finite volume
with periodic boundary conditions~\cite{Fodor:2012td}, and recently in a box with
twisted boundary conditions~\cite{Ramos:2013gda}. Here we are interested in the 
gradient flow (GF) coupling in a finite volume with Schr\"odinger functional (SF)
boundary conditions, which is defined as~\cite{Fritzsch:2013je}
\begin{equation}
\label{eq:GFcoupling}
\overline{g}^2_{\rm GF}(L) \equiv 
\mathcal{N}^{-1} \langle t^2E(t,x_0)\rangle|_{t=c^2L^2/8},\quad T=L,\quad x_0=T/2,
\end{equation}
where $c$ is a constant that relates the flow time to the spatial extent $L$,
and $\mathcal{N}$ ensures the correct normalization 
$\overline{g}^2_{\rm GF}=g_{\overline{\rm MS}}^2 + O(g_{\overline{\rm MS}}^4)$.
The specific scheme in (\ref{eq:GFcoupling}) (and so the constant $\mathcal{N}$)
in fact depends on the Dirichlet boundary conditions at $x_0=0,\,T$, and the values
of $c$, $x_0$, and $T/L$ one considers.\footnote{Note that SF boundary conditions
break translational invariance in time. The flow energy density $E$ thus depends 
explicitly on the time coordinate $x_0$.} The SF boundary conditions can be chosen
such that there is a unique global minimum of the action (up to gauge transformations)
around which the perturbative expansion of the coupling is easy to set 
up~\cite{Fritzsch:2013je,Luscher:1992an}. This is indeed crucial for the application
we are going to discuss.

In these Proceedings we study the GF coupling in the SF using lattice perturbation
theory. From perturbation theory one can extract valuable information for a
non-perturbative determination of the running coupling. First of all, the matching
to other schemes is generally done at high energies using perturbation theory. 
Secondly, cutoff effects in the step-scaling function can be determined perturbatively
and be used to improve non-perturbative data~\cite{deDivitiis:1994yz}. In order to make
these determinations more effective, however, perturbation theory needs to be pushed
beyond the 1-loop order, which is in most cases technically involved. Although codes
have been developed for the automation of such calculations~\cite{Hesse:2012hb},
the inclusion of the gradient flow and the need for high order contributions
complicate things further, rendering these computations rather challenging. For this 
reason we rely on Numerical Stochastic Perturbation Theory (NSPT) (see~\cite{DiRenzo:2004ge}
for an introduction). Due to the similarities between the Langevin and gradient 
flow equations, NSPT is a natural framework for a perturbative numerical solution
of the flow. In the next section we start with a reminder of the lattice setup as
discussed in~\cite{Fritzsch:2013je}. We then recall the basics of NSPT and explain
how the gradient flow equation can be solved in this framework. Finally, we present
our results for the flow energy density $E$ to 3-loops in perturbation theory which
gives direct access to $\overline{g}^2_{\rm GF}$ at 2-loops.

\section{The gradient flow coupling on the lattice}
\label{sec:lat}

The gradient flow equation (\ref{eq:GF}) can be studied on the lattice by introducing
the field $V_\mu(t,x)$ (also known as ``Wilson flow''~\cite{Luscher:2010iy}) 
defined by the equation
\begin{equation}
 \label{eq:GFlat}
  {\partial_t}{V}_\mu(t,x) = 
  -\big\{g_0^2\nabla_{x\mu}S_W({V}(t))\big\}{V}_\mu(t,x),\quad
   V_\mu(0,x) = U_\mu(x),
\end{equation}
where $\nabla_{x\mu}$ is the Lie-derivative with respect to $V_\mu(t,x)$, and $S_W$
is the Wilson plaquette action. Analogously, we choose the Wilson plaquette action
also for the gauge field $U_\mu(x)$. Following~\cite{Fritzsch:2013je}, we then 
consider an $(L/a)^4$ lattice and impose, for all values $t\geq 0$ of the flow time, 
the SF boundary conditions:
\begin{equation}
\label{eq:bc}
     V_\mu(t,x+\hat{k}L)= V_\mu(t,x),\quad V_k(t,x)|_{x_0 = 0,\,L} = \mathds{1}.
\end{equation}
Given these boundary conditions, 
the action $S_W$ has a unique global minimum (up to gauge transformations)
corresponding to the field configuration $\mathcal{V}_\mu(x)=\mathds{1}$, $\forall x,\mu$.
Finally, we define the energy density $E$ on the lattice through the continuum 
formula (\ref{eq:Eoft}) and the lattice definition of the field strength tensor 
$G_{\mu\nu}$ given in~\cite{Luscher:2010iy}. In fact, as noticed in~\cite{Fritzsch:2013je},
a renormalized coupling can be defined as in (\ref{eq:GFcoupling}) considering 
only the \emph{spatial} or \emph{temporal} components of the strength tensor in 
the expression for $E$. Later, we will refer to the corresponding contributions 
as $E_s$ and $E_t$, respectively.

\section{A numerical perturbative solution for the gradient flow}

The idea of NSPT is to solve the equations of stochastic perturbation
theory numerically~\cite{DiRenzo:1994av}. One starts from the stochastic 
quantization of the lattice theory, where the fundamental fields evolve in
the extra coordinate $t_s$ (known as \emph{stochastic time}) according
to the Langevin equation
\begin{equation} 
\label{eq:Langevin}
  {\partial_{t_s}} U_\mu(t_s,x) = 
  -\big\{\nabla_{x\mu}S_W(U(t_s)) + \eta_\mu(t_s,x)\big\}U_\mu(t_s,x),
\end{equation}
where $\eta_\mu(t_s,x)$ is a Gaussian distributed noise field.
The Langevin equation is then discretized in the stochastic time, $t_s=n\varepsilon_s$,
and a general solution is obtained according to a given integration scheme
e.g. Runge-Kutta. Last, introducing in the solution the perturbative
expansion
\begin{equation}
  \label{eq:UPT}
  U_\mu(t_s,x) \to \sum_{k} g_0^k\, U^{(k)}_\mu(t_s,x),\quad 
  U^{(0)}_\mu(t_s,x)\equiv\mathcal{V}_\mu(x),
\end{equation}
where $\mathcal{V}$ is the field configuration around which we expand, 
the result is a hierarchy of equations that can be truncated 
consistently\footnote{Note that the stochastic time has to be 
rescaled as $t_s\to g_0^2\,t_s$ in order to make this expansion consistent.} 
and solved numerically. In the limit of $t_s\to\infty$ the noise 
average of any observable $\mathcal{O}$ evaluated on the perturbative
solution (\ref{eq:UPT}) is then expected to converge order-by-order in perturbation
theory to the corresponding expectation value of the lattice theory, i.e.
\begin{equation}
\label{eq:NSPT}
\lim_{t_s\to\infty}
\big\langle{\mathcal{O}\big[\sum_k g_0^k\,U^{(k)}(t_s)\big]}\big\rangle_\eta
=\lim_{t_s\to\infty}\sum_k\,g_0^k\,\langle{\mathcal{O}}^{(k)}(t_s)\rangle_\eta
=\sum_k\,g_0^k\,\mathcal{O}_{k},
\end{equation}
where $\langle\cdots\rangle_\eta$ is the average over the noise field distribution, 
and $\mathcal{O}_k$ is the $k$-th order coefficient of the perturbative expansion of 
$\langle\mathcal{O}\rangle$. The gradient flow can now be included as follows. 
Compared to the Langevin equation, there are two important differences to consider: 
in the flow equation the noise term is absent, and the initial distribution
of the gauge field $U_\mu(x)$ (here given by a perturbative expansion of the Langevin
equation) has to be taken into account (cf. (\ref{eq:GFlat})). To this end, we 
introduce in a discrete solution of the flow equation
\begin{equation}
\label{eq:VPT}
  {V}_\mu(t_s;t,x) \to 
  \sum_{k} g_0^k\,{V}^{(k)}_\mu(t_s;t,x),\quad
  V^{(k)}_\mu(t_s;0,x)=U^{(k)}_\mu(t_s,x),\quad\forall k,
\end{equation}
where $t=n\varepsilon$, and $V_\mu(t_s;t,x)$ inherits the $t_s$ dependence from the initial condition.
The result of the expansion is the same hierarchy of equations as for the Langevin
equation, except that the noise field is set to zero. For a given initial gauge
field configuration, the flow equation can thus be integrated numerically 
up to the desired flow time, order-by-order, using the same techniques. In particular,
note that the flow field (\ref{eq:VPT}) is a function of the gauge field (\ref{eq:UPT})
through the initial condition. The perturbative expansion of any flow observable 
is then obtained as in (\ref{eq:NSPT}).

To conclude, both perturbative solutions of the Langevin and flow equations are  
derived from a discrete approximation of these equations. Results have then to 
be extrapolated for $\varepsilon_s,\varepsilon\to0$ in order to eliminate the effects
of the discretization. In addition, in NSPT stochastic gauge fixing is fundamental
to obtain sensible results~\cite{DiRenzo:2004ge}. We refer however to~\cite{Hesse:2013lat}
for a detailed discussion of stochastic gauge fixing in NSPT for SF schemes.

\section{Results}

\subsection{Determination of $\mathcal{N}$ and comparison with analytical results}

\begin{figure}[htb]
  \centering
  {\includegraphics[width=.48\textwidth]{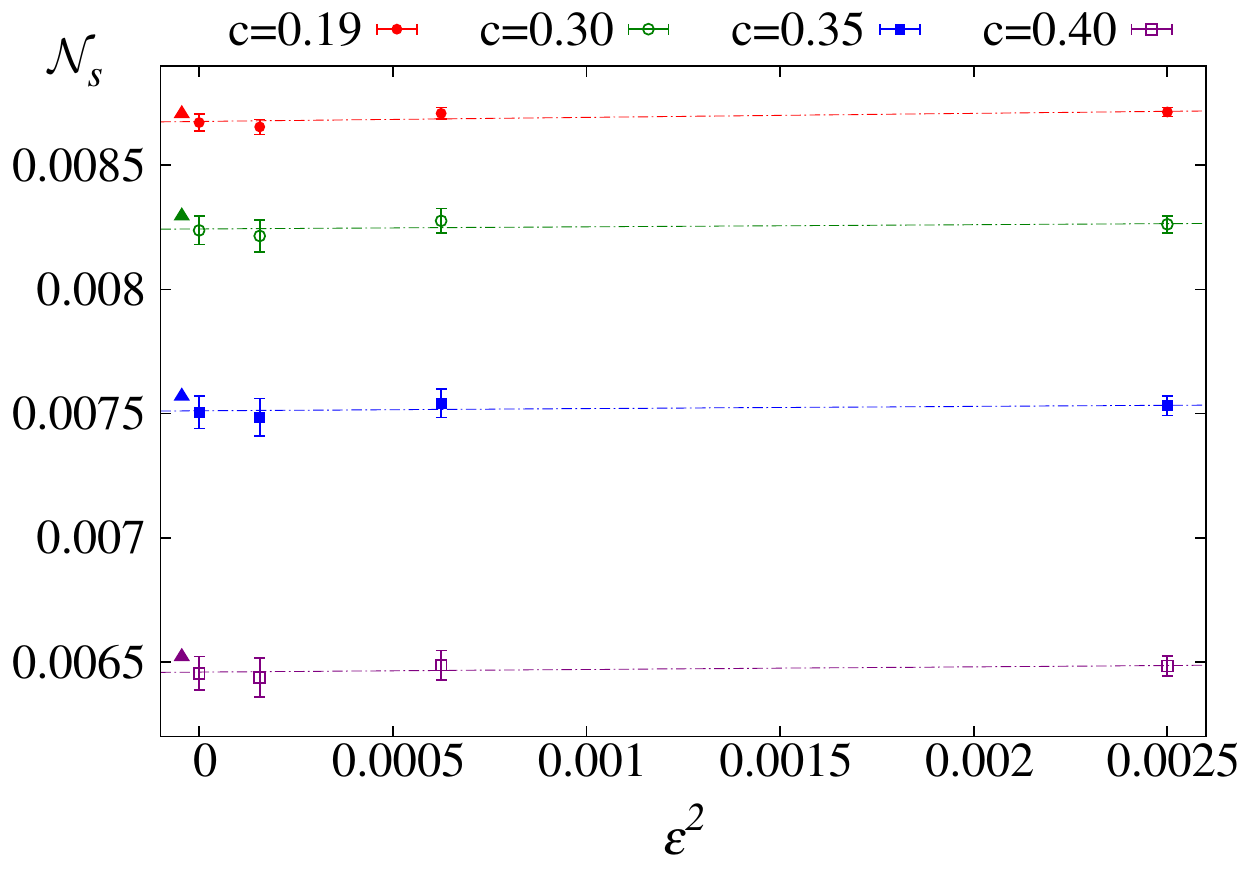}} \quad
  {\includegraphics[width=.48\textwidth]{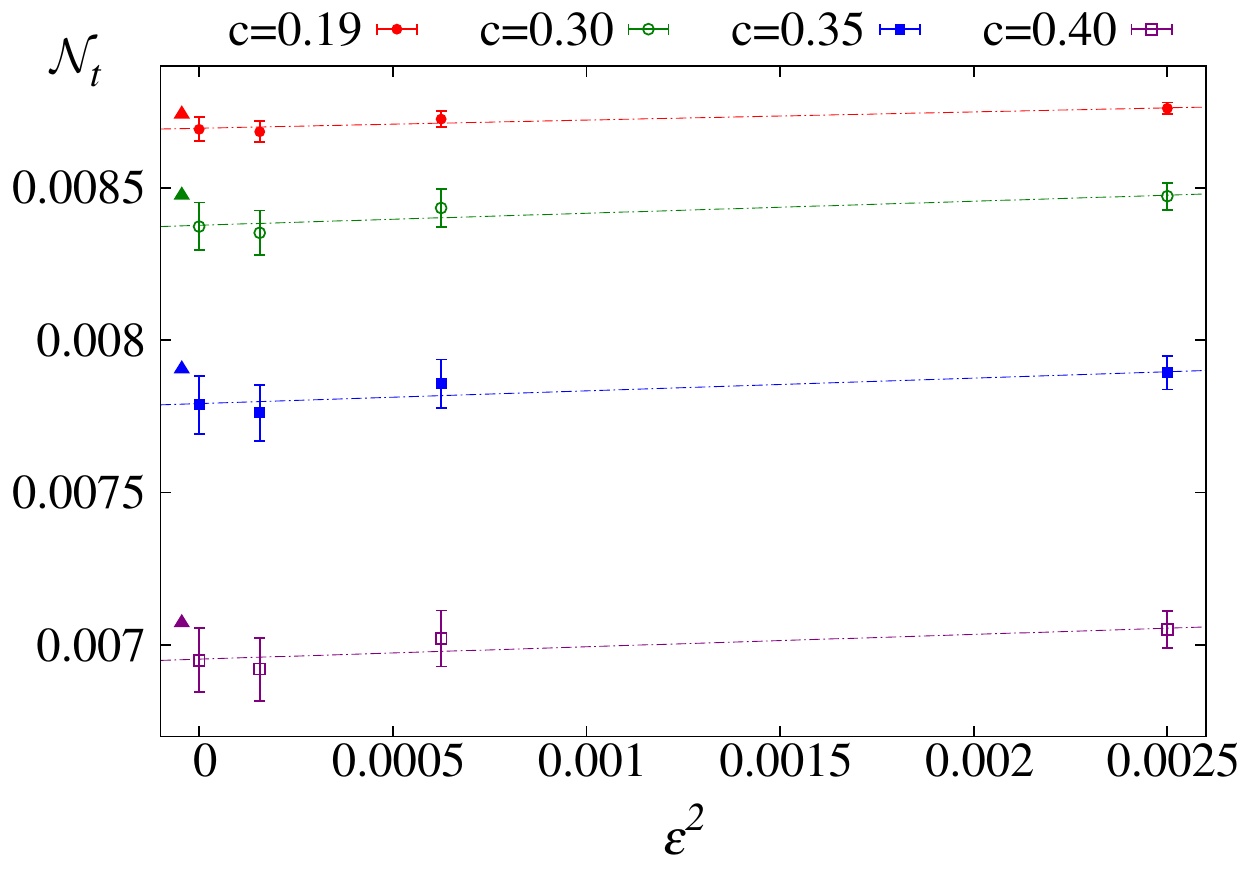}}
  \caption{Extrapolations of $\mathcal{N}$ to zero step-size for $L/a=12$
  and several schemes $c$. \emph{Left:} Values for the spatial contribution $\mathcal{N}_{s}$.
  \emph{Right:} Values for the temporal contribution $\mathcal{N}_{t}$.
  Analytical results are depicted as triangles.}
  \label{fig:LO}
\end{figure}

As discussed before, a renormalized coupling can be defined considering
the separate contributions
\begin{equation}
\label{eq:EoftPT}
{\mathcal{E}}_i \equiv \langle t^2 E_i(t,L/2) \rangle  
= \mathcal{N}_i\,g_0^2 + {\mathcal{E}}_i^{(1)}\,g_0^4+ 
 {\mathcal{E}}_i^{(2)}\,g_0^6 + \ldots,\quad  i=s,\,t,
\end{equation}
where the lowest-order coefficient $\mathcal{N}$ enters as part of the definition of the coupling
(we leave out the subscript $i$ when a statement holds for both $s$ and $t$).
This term has been computed in~\cite{Fritzsch:2013je} for the lattice setup we are
considering. As a first result we reproduced the values of $\mathcal{N}$
for lattices up to $L/a=12$, and different choices of $c$ (cf.~\ref{eq:GFcoupling}). 
Some examples are collected in fig. \ref{fig:LO}. The data are extrapolated in $\epsilon^2$
since a second-order integrator was used for both the Langevin and flow equation 
and we took $\varepsilon_s=\varepsilon$.
In particular, we considered a linear fit to all data points and a constant fit to the
data omitting $\varepsilon = 0.05$, and estimated systematic effects as the difference
between the two fits. The extrapolated values agree within errors with the 
analytical results.

\subsection{Determination of $\mathcal{E}^{(1)}$ and comparison with Monte Carlo data}

Next, we consider the NLO contribution given by $\mathcal{E}^{(1)}$.
For this quantity no independent determinations are available for direct comparison.
Hence, as a crosscheck on our results, we  obtained an estimation of 
$\mathcal{E}^{(1)}$ from Monte Carlo (MC) simulations. We performed
pure SU(3) gauge simulations at 12 values of $\beta=6/g_0^2\in[50,1200]$ at fixed 
$L/a$, and we extracted $\mathcal{E}^{(1)}$ from a fit of $\mathcal{E}$ to (\ref{eq:EoftPT})
fixing $\mathcal{N}$ to its analytical value. The results we obtained for
$L/a=8$ are summarized in Table \ref{tab:NSPTvsMC}. The agreement between NSPT 
and MC simulations is generally good (max deviation $\sim 2.6\,\sigma$), and we 
note NSPT to be systematically more precise at a comparable computational cost.
\begin{table}[hbt]
  \centering
  \scalebox{.90}{
  \begin{tabular}{llllll}
   \toprule
   &\multicolumn{2}{c}{$\mathcal{E}^{(1)}_s$} & 
   &\multicolumn{2}{c}{$\mathcal{E}^{(1)}_t$} \\
   \cmidrule{2-3}\cmidrule{5-6}
   \multicolumn{1}{c}{$c$} & \multicolumn{1}{c}{MC} & \multicolumn{1}{c}{NSPT} & & 
   \multicolumn{1}{c}{MC} & \multicolumn{1}{c}{NSPT}\\\midrule
   0.19 & 0.00478~(\phantom{0}9)  & 0.00463~(2)  && 0.00486~(\phantom{0}9)  & 0.00461~(2) \\
   0.30 & 0.00552~(15)  & 0.00546~(5)  && 0.00557~(17)  & 0.00541~(5) \\
   0.40 & 0.00483~(18)  & 0.00478~(6)  && 0.00479~(22)  & 0.00472~(7) \\
   0.50 & 0.00355~(14)  & 0.00349~(6)  && 0.00351~(21)  & 0.00350~(6) \\
   \bottomrule   
  \end{tabular}}
  \caption{Comparison between Monte Carlo and NSPT results for $\mathcal{E}^{(1)}$
  for an $L/a=8$ lattice and different values of $c$. 
  The spatial $\mathcal{E}^{(1)}_s$ and the temporal contribution $\mathcal{E}^{(1)}_t$ are shown.}
  \label{tab:NSPTvsMC}
\end{table}  

\subsection{Determination of $\mathcal{E}^{(2)}$}

\begin{figure}[htb]
  \centering
  {\includegraphics[width=.48\textwidth]{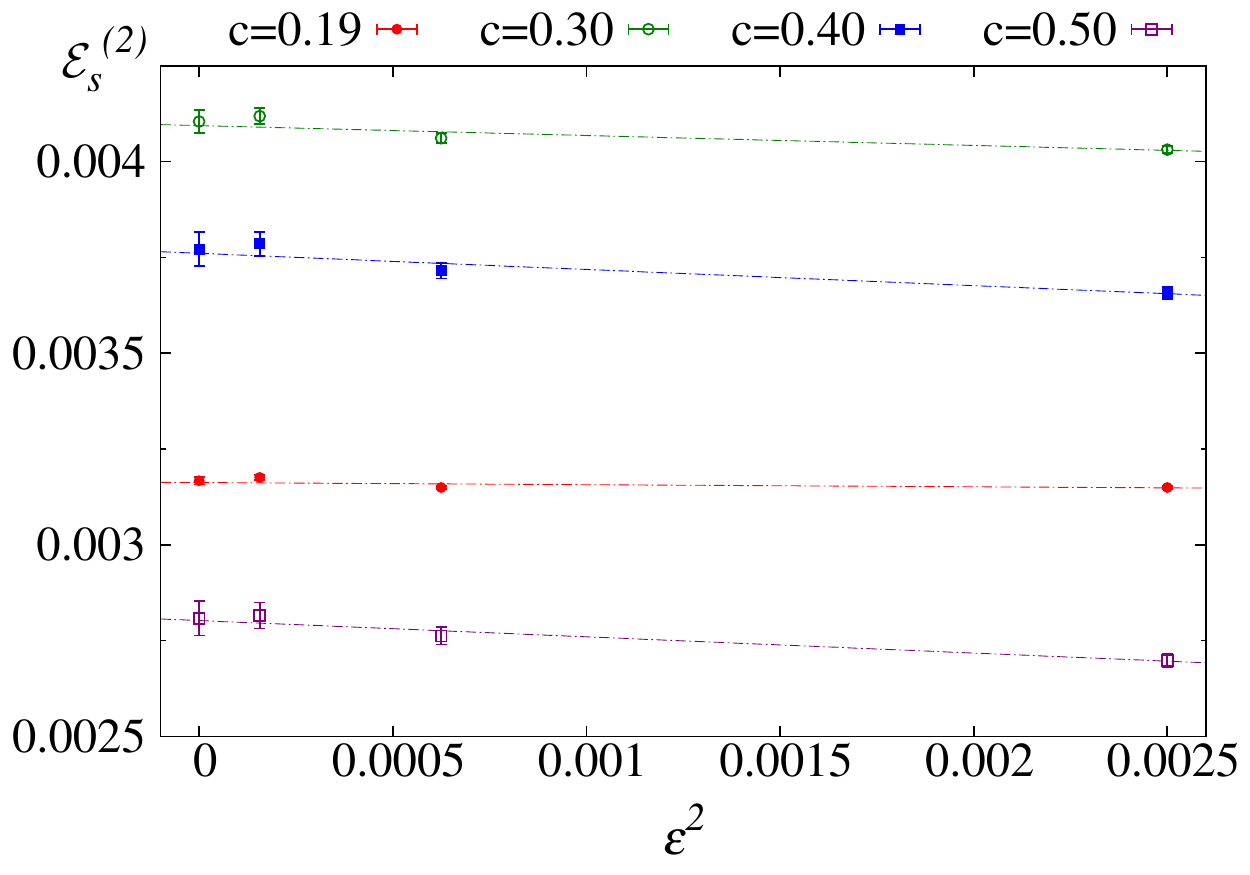}} \quad
  {\includegraphics[width=.48\textwidth]{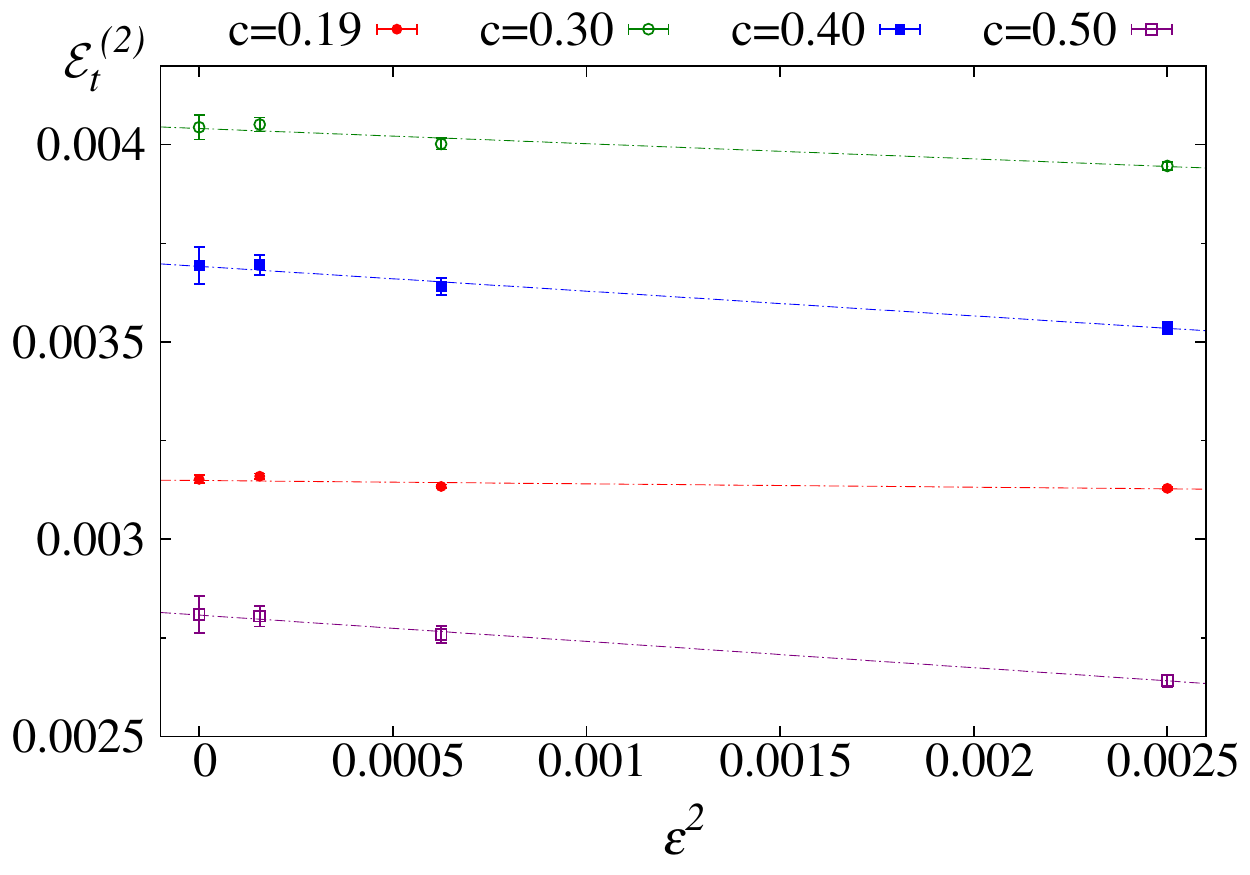}} \\
  \caption{Extrapolations of $\mathcal{E}^{(2)}$ to zero step-size for $L/a=8$
  and several schemes $c$. \emph{Left:} Values for the spatial contribution 
  $\mathcal{E}^{(2)}_{s}$. \emph{Right:} Values for the temporal contribution 
  $\mathcal{E}^{(2)}_{t}$.}
  \label{fig:NNLO}
\end{figure}

The highest order we have computed is the NNLO contribution $\mathcal{E}^{(2)}$. 
An example of the $\varepsilon\to0$ extrapolation for an $L/a=8$ lattice is shown 
in fig. \ref{fig:NNLO}. The results of the extrapolations have been compared
with the ones obtained  by MC simulations. Although agreement was found, the errors
on the extracted values for $\mathcal{E}^{(2)}$ from MC simulations were quite large
(about 20-50\%), thus providing not such a stringent constraint on our determination.
Smaller values of $\beta$ have probably to be considered in order to resolve this 
contribution from the MC data. The results obtained from NSPT however have a
good overall precision of $\sim1\%$ for this small lattice size. 

\subsection{Relative variance and autocorrelations}

\begin{figure}[hbt]
  \centering
  {\includegraphics[width=.48\textwidth]{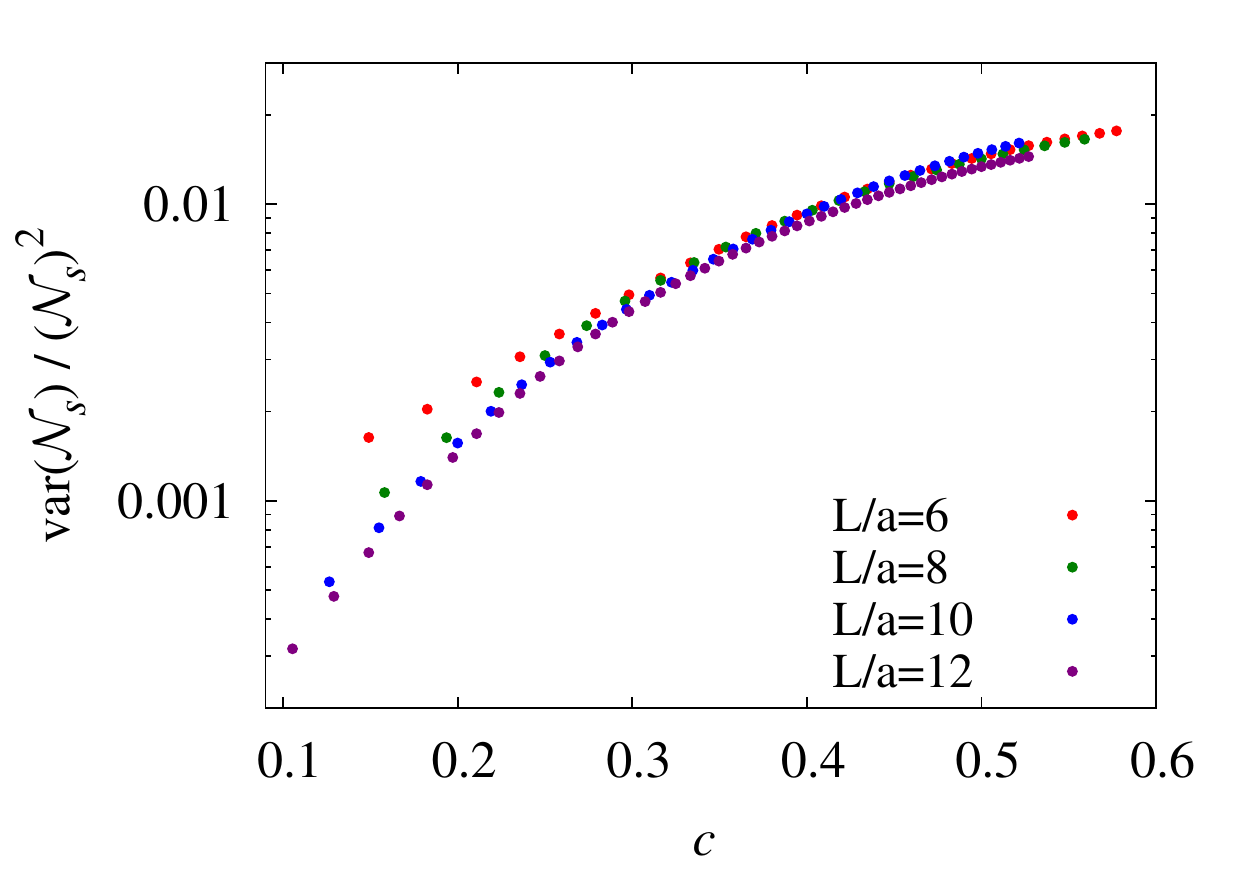}} \quad
  {\includegraphics[width=.48\textwidth]{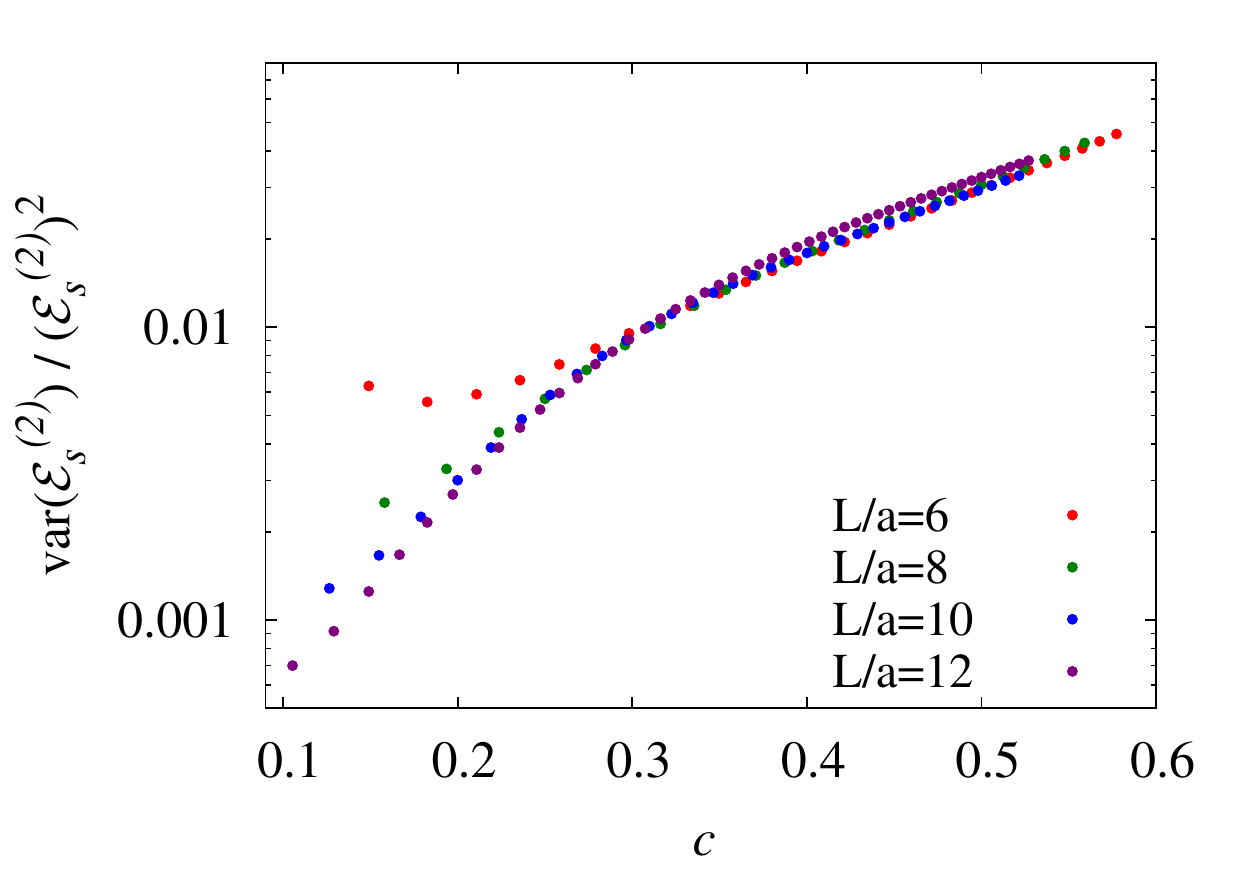}}
  \caption{Relative variance as a function of $c$ for different $L/a$ 
  for computations at $\varepsilon = 0.0125$.}
  \label{fig:var}
\end{figure}
Finally, we present some results on the variance of $\mathcal{E}_s$
(analogous results hold for $\mathcal{E}_t$). More precisely, in fig. \ref{fig:var}
we show the relative variance of $\mathcal{N}_s$ and $\mathcal{E}^{(2)}_s$ as a function
of $c$ for different lattice sizes $L/a$. Here we consider only computations with 
$\varepsilon=0.0125$ which are the most expensive ones we performed. The first 
observation is that the relative variance is rather independent of $L/a$ for schemes 
with $c \gtrsim 0.3$. Secondly, higher order terms have relative variance comparable  
to the lowest ones, indicating that a similar level of precision can be reached. 
Note in particular that the results in fig. \ref{fig:var} give direct information 
on the number of independent measurements necessary to obtain a certain precision
on the result. For example, to compute $\mathcal{N}_s$ for $c\sim 0.3$ at the 0.1\%
level we need $\sim 5000$ independent measurements, for $\mathcal{E}_s^{(2)}$ instead
$\sim 9000$ are necessary. The level of precision on $\overline{g}^2_{\rm GF}$ one
can achieve then seems relatively good with moderate statistics. To conclude, one
has to consider that in order to obtain $\mathcal{E}^{(k)}$ for a given scheme $c$
at a certain level of precision, the cost of the computation scales $\propto (L/a)^6$.
Apart from the volume factor $\propto (L/a)^{4}$, the cost of integrating the flow
increases $\propto (L/a)^2$ for fixed $c$ (cf. (\ref{eq:GFcoupling})). Autocorrelations
also increase $\propto (L/a)^2$ as expected for the Langevin dynamics. The measurement
frequency of the flow along the stochastic time then can be scaled $\propto (L/a)^{-2}$
without significant loss in the statistical precision of the determination.

\section{Conclusions \& outlook}

We studied the flow energy density $E$ in the SF at 3-loops
in perturbation theory for the pure SU(3) gauge theory. We reproduced the values
for $\mathcal{N}$ in~\cite{Fritzsch:2013je} and compared higher orders to 
MC results finding good agreement. Our study shows that a precise determination 
of $\overline{g}^2_{\rm GF}$ using  NSPT is feasible with moderate statistics. 
As a next step we plan to investigate the GF coupling in QCD. Larger lattice sizes
will be considered in order to extrapolate the matching coefficients to other schemes,
and the corresponding cutoff effects in the step-scaling function will be determined.

\section*{Acknowledgments}

The authors are grateful to M. L\"uscher, S. Sch\"afer, S. Sint, and R. Sommer for 
their valuable comments. We thank A. Ramos for kindly sharing material before 
publication, and F. Di Renzo and L. Scorzato for discussions. We also thank
M. Bruno, T. Harris, S. Lottini, and P. Vilaseca for kindly providing comments
on this manuscript. M.D.B. is supported by the Irish Research Council. 
D.H. acknowledges support by StrongNet. The code for MC simulations is 
based on the \texttt{MILC} package. ICHEC, TCHPC, and AuroraScience are 
acknowledged for the allocated resources.

\bibliographystyle{JHEP}
\bibliography{bibliography}

\end{document}